\documentclass[12pt]{article}
\usepackage{amsthm,amsmath,amsfonts,amssymb}
\usepackage{amsmath,amsfonts,amsthm}
\newcommand{\R}{{\mathord{\mathbb R}}}
\newcommand{\Z}{{\mathord{\mathbb Z}}}

\newcommand{\C}{{\mathord{\mathbb C}}}

\newcommand{\HH}{\mathcal{H}}

\newcommand{\FF}{\mathcal{F}}

\newcommand{\hh}{\mathfrak{h}}

\newcommand{\ran}{{\rm Ran}}

\newcommand{\ben}{\begin{displaymath}}
\newcommand{\een}{\end{displaymath}}
\newcommand{\beqn}{\begin{equation}}
\newcommand{\eeqn}{\end{equation}}
\newcommand{\beqna}{\begin{eqnarray*}}
\newcommand{\eeqna}{\end{eqnarray*}}

\def\inf{{\rm inf}\,}

\newcommand{\sfrac}[2]{\textrm{\footnotesize $\frac{#1}{#2}$}}

\newcommand{\restricted}{|\grave{}\,}

\newtheorem{lemma}{Lemma}
\newtheorem{theorem}[lemma]{Theorem}

\newtheorem{proposition}[lemma]{Proposition}

\title{Absence of Ground States for a Class of
Translation Invariant Models of Non-relativistic QED}

\date{ }

\author{D. Hasler, I. Herbst}

\begin{document}

\maketitle

\begin{abstract}
We consider a class of translation invariant models of
non-relativistic QED with net charge. Under certain natural
assumptions we prove that ground states do not exist in the Fock
space.
\end{abstract}

\section{Introduction}

Over the years there has been much interest in trying to develop an
appropriate mathematical framework to describe the interaction of
charged particles with the quantized electromagnetic field.  Here we
only cite \cite{SPOHN04} and references given therein but later we
briefly mention other work.  Of course relativistic quantum
electrodynamics (QED) is a very successful theory but has not been
shown to provide a Hilbert space framework for describing the states
of charged particles interacting with photons.  In spite of this
there are certainly prescriptions for getting correct answers to the
``right'' questions \cite{IZ80}.

One of the first questions which arises is perhaps the most
elementary:  Are there ``dressed one-electron states'' of fixed
momentum which are eigenstates of the appropriate Hamiltonian. These
states should of course have an adhering photon cloud.   In
\cite{FK} Faddeev and  Kulish gave a suggestion as to what form such
states should take.  The Fadeev-Kulish states do not live in Fock
space because of the nature of the photon cloud.  At this time,
however, we are far from understanding the mathematics of
relativistic QED.

In order to understand the infrared problem in a simpler model,
Fr\"ohlich \cite{F73,F74}, studied the massless Nelson model.  This
is a model of a non-relativistic particle interacting with a scalar
massless bose field ("photon" field).  Among other results, in
\cite{F73} he outlined a construction of asymptotic dressed one
particle states (with a low energy photon cloud).  Recently, Pizzo
\cite{P05} has taken Fr\"ohlich's outline, added some important
ingredients, and rigourously constructed a Hilbert space of
asymptotic dressed one-particle states (with certain smallness
assumptions on particle velocity and on various parameter values).

In recent years the more realistic model of non-relativistic QED has
been studied by many authors, see for example \cite{SPOHN04} and
references given therein. This model suffers from various
difficulties but it is hoped that it may serve as a reasonably
realistic model for low energies, and a testing ground for
understanding the infrared problem.  One of the main difficulties is
that this model is neither Galilean nor Poincar\'{e} covariant.  The
charged particles are treated non-relativistically while the photons
are relativistic. There remains an ultra-violet cutoff in the photon
field to produce a well defined theory, but the theory is well
defined without an infrared cutoff. More recently, Chen and
Fr\"ohlich \cite{CF06} have also outlined the construction of
asymptotic dressed one-particle states in non-relativistic QED,
partly relying on some of the ideas in \cite{F73,P05}.

In this work we define our Hamiltonians on the Hilbert space
consisting of the Fock space for photons tensored with the usual
Hilbert space for the non-relativistic charged particles. We
consider a class of translation invariant models of non-relativistic
QED having a total net charge. The generator of translations defines
the operator of total momentum. Translation invariance implies that
the Hamiltonian commutes with this operator. We can thus  restrict
the Hamiltonian to any subspace of fixed total momentum $\xi$. This
restricted Hamiltonian is denoted by $H(\xi)$. For any momentum
$\xi$, $H(\xi)$ is bounded from below. We denote the infimum of its
spectrum by $E(\xi)$. One can easily show the the function
$E(\cdot)$ is almost everywhere differentiable. In this paper we
show that for momenta $\xi$ at which $E(\cdot)$ has a non-vanishing
derivative, $H(\xi)$ does not admit a ground state. We do not impose
an infrared cutoff, which in fact is the reason for the absence of
ground states. The coupling constant is arbitrary, but nonzero.

First we consider an electron (with spin $1/2$) coupled to the
quantized electromagnetic field. We show that  for any value of the
coupling constant $H(\cdot)$ does not admit a ground state at points
where $E(\cdot)$ has a non-vanishing derivative. This model has been
previously investigated in \cite{C01,BCFS05,CF06}. There it was
shown that for small values of the coupling constant, $E(\cdot)$
 has a non-vanishing derivative for all nonzero $\xi$
with $|\xi|< \xi_0$, where $\xi_0$ is some explicit positive number.
Furthermore, for small coupling it was shown that $H(0)$ does have a
ground state. Moreover, for small coupling and nonzero $\xi$, with
$|\xi|<\xi_0$, it was shown that an infrared regularized Hamiltonian
does have a ground state. As the infrared regularization is removed
this ground state does not converge in Fock space, however it can be
shown that it does converge as a linear functional on some operator
algebra, \cite{F73,CF06}.

The model is introduced and the result is stated in Section
\ref{sec:eldef}. The proof of the result is presented in Section
\ref{sec:elpro}. Although on the basis of the work cited above, our
result is expected, we have not found a proof in the literature.

We then generalize the above result to a positive ion. More
specifically, we consider a spinless nucleus with nuclear charge
$Ze$ and $N$ electrons each with charge $-e$ where the interaction
between the particles includes the Coulomb potential. If $Z \neq N$,
we show that $H(\cdot)$ does not admit a ground state at points
where $E(\cdot)$ has a non-vanishing derivative. This model has been
recently investigated in \cite{LMS06,AGG06}, where it was shown that
under natural assumptions $H(\xi)$ does have a ground state provided
$N=Z$. It was known previously that if the nucleus has infinite
mass, then the relevant Hamiltonian does have a ground state if $Z
\geq N$, \cite{BFS99,GLL01}. In contrast to our result,  Coulomb
systems without coupling to the quantized electromagnetic field  do
have positive ions,  with fixed nonzero total momentum. In Section
\ref{sec:eldef} we introduce the model describing an ion  and state
the result. Its proof is presented in Section \ref{sec:elpro}.
Although perhaps surprising, the intuition for our result comes from
the fact that from a distance, a charged bound state looks like a
point particle.

In order to show that the physical properties of the theory do not
depend on an ultraviolet cutoff, small coupling results where the
coupling depends on the ultraviolet cutoff are typically not
sufficient. The proof of our result employs the so called
pull-through formula. In order to deal with arbitrary values of the
coupling constant we have to restrict our analysis to a subset of
momentum space. This however is sufficient to rule out the existence
of a ground state. In the next section  we introduce the Fock space
of photons.

\section{Fock Space of  Photons}

\label{sec:nonqed}

 The degrees of freedom of the photons are
described by a symmetric Fock space, introduced as follows. Let
$$
\hh := L^2(\Z_2  \times \R^3 ) \cong L^2(\R^3 ; \C^2 )  \;
$$
denote the Hilbert space  of a transversally polarized photon. The
variable $\underline{k} = (\lambda,k) \in  \Z_2  \times \R^3$ consists
of the wave vector $k$ or momentum of the particle and  $\lambda$
describing the polarization. The symmetric Fock space, $\FF$, over
$\hh$ is defined by
$$
\FF = \C \oplus \bigoplus_{n=1}^\infty S_n( \hh^{\otimes n} ) \; ,
$$
where $S_n$ denotes the orthogonal projection onto the subspace of
totally symmetric tensors. The vacuum  is the vector $\Omega :=
(1,0,0, ... ) \in \FF$. The  vector $\psi \in \FF$ can be identified
with sequences $(\psi_n)_{n=0}^\infty$ of $n$-photon wave functions,
$\psi_n(\underline{k}_1, ... , \underline{k}_n) \in L^2((
\Z_2 \times \R^3)^n)$, which for $n \geq 1$ are totally symmetric in their $n$
arguments. The Fock space inherits a scalar product from $\hh$,
explicitly
$$
( \psi , \varphi )_\FF =  \overline{\psi}_0 \varphi_0 +
\sum_{n=1}^\infty \int \overline{\psi}_n(\underline{k}_1, ... ,
\underline{k}_n) \varphi_n(\underline{k}_1, ... , \underline{k}_n )
d \underline{k}_1 ...  d \underline{k}_n \; ,
$$
where we used the abbreviation $\int d \underline{k} =
\sum_{\lambda=1,2} \int dk$. The number operator $N$ is defined by
$(N \psi)_n = n \psi_n$. It is self-adjoint on the domain $D(N) :=
\{ \psi \in \FF | N\psi \in \FF\}$. For each function $f \in \hh$
one associates an annihilation operator $a(f)$ as follows.  For a
vector $\psi \in \FF$ we define
$$
(a(f) \psi )_n(\underline{k}_1, ... , \underline{k}_n ) =
(n+1)^{1/2} \int  \overline{f}(\underline{k})
\psi_{n+1}(\underline{k}, \underline{k}_1, ... , \underline{k}_n ) d
\underline{k} \quad , quad \forall \ n \geq 0 \;  .
$$
The domain of $a(f)$ is the set of all $\psi$ such that $a(f)\psi
\in \FF$. Note that $a(f) \Omega = 0$. The creation operator
$a^*(f)$ is defined to be the adjoint of $a(f)$. Note that
 $a(f)$ is anti-linear, and $a^*(f)$ is linear in $f$. They are well
 known to satisfy the canonical commutation relations
$$
[ a^*(f) , a^*(g) ] = 0 \quad , \quad [a(f) , a(g) ] = 0 \quad ,
\quad [a(f) , a^*(g) ] =  (f , g) \; .
$$
where $f , g \in L^2( \Z_2 \times \R^3 )$ and  $(f , g)$ denotes the
inner product of $L^2( \Z_2 \times \R^3 )$. Since $a(f)$ is
anti-linear, and $a^*(f)$ is linear in $f$, we will write
$$
a(f) = \int     \overline{f}(\underline{k})   a_{\underline{k}}
d\underline{k} \quad , \quad a^*(f) = \int      f(\underline{k})
a^*_{\underline{k}} d\underline{k} \; ,
$$
where the right hand side is merely a different notation for the
expression on the left. For a function $f \in L^2(\R^3)$ and
$\lambda=1,2$, we will write $a_\lambda(f) := a(f_\lambda)$ and
$a^*_\lambda(f) := a^*(f_\lambda)$, where $f_\lambda \in \hh$ is the
function defined by $f_\lambda(\mu,k) := f(k) \delta_{\lambda,\mu}$.
The field energy operator denoted by $H_f$ is given by
$$
(H_f \psi)_n(\underline{k}_1, ... \underline{k}_n) = \left(
\sum_{i=1}^n |k_i| \right) \psi_n(\underline{k}_1, ...
\underline{k}_n) \; .
$$
It is self-adjoint on its natural domain
$D(H_f) := \{ \psi \in \FF | H_f \psi \in \FF \}$.
The operator of momentum $P_f$ is given by
$$
(P_f \psi)_n(\underline{k}_1, ... \underline{k}_n) = \left(
\sum_{i=1}^n k_i \right) \psi_n(\underline{k}_1, ...
\underline{k}_n) \; .
$$
Its components $(P_f)_j$ are each self-adjoint on the domain
$D((P_f)_j) := \{ \psi \in \FF | (P_f)_j \psi \in \FF\}$. In this
paper we will adapt the notation that $| \cdot |$ denotes the
standard norm in $\R, \R^3, \C$, or $\C^2$.

\section{The Electron: Model and Statement of Result}

\label{sec:eldef}

 At first  we consider a single free electron
interacting with the quantized electromagnetic field. The Hilbert
space describing the system composed of an electron and the
quantized field is
$$
\HH = L^2( \R^3 ; \C^2)  \otimes \FF \; .
$$
The Hamiltonian is
$$
H =   \left\{  \sigma \cdot ( p  + e A (x) ) \right\} ^2 + H_f \; ,
$$
where
\begin{equation} \label{eq:44}
A(x) = \sum_{\lambda=1,2} \int \frac{\rho(k)}{\sqrt{2|k|}} \left(
a_{\lambda,k} e^{ik \cdot x} \varepsilon_{\lambda,k}  +
a_{\lambda,k}^* e^{-ik \cdot x} \varepsilon_{\lambda,k}  \right) dk
\; ,
\end{equation}
where the $\varepsilon_{\lambda,k} \in \R^3$ are vectors, depending
measurably on $\widehat{k} = k /|k|$,  such that $(k/|k|,
\varepsilon_{1,k}, \varepsilon_{2,k})$ forms an orthonormal basis;
and  $\sigma = (\sigma_1, \sigma_2 , \sigma_3)$, where $\sigma_i$
denotes the $i$-th Pauli matrix:
$$
\sigma_1 = \left( \begin{array}{cc} 0 & 1 \\ 1 & 0   \end{array}
\right) \ \ , \ \ \sigma_2 = \left(
\begin{array}{cc}  0 & - i \\ i & 0  \end{array} \right) \ \ , \ \ \sigma_3 =
\left( \begin{array}{cc} 1 & 0 \\ 0 & -1    \end{array} \right) \; .
$$
By $x$ we denote the position of the electron and its canonically conjugate
momentum by $p = - i \nabla_x$.
We have introduced the function
$$
\rho(k) = \frac{1}{(2 \pi)^{3/2}} \chi_\Lambda(|k|) \; ,
$$
where $\chi_{\Lambda}$ is the characteristic function of the set
$[0,\Lambda]$. Since we are interested in the infrared problem we
fix the ultraviolet cutoff $0 < \Lambda < \infty$. The Pauli
matrices satisfy the commutation relations $[\sigma_1 ,\sigma_2] = 2
i \sigma_3$ and cyclic permutations thereof. Using these commutation
relations, we can write the Hamiltonian as
$$
H = ( p + e A(x) )^2 +  e \sigma \cdot B(x) +  H_f \; ,
$$
where
$$
B(x) = (\nabla \wedge A)(x) =   \sum_{\lambda=1,2} \int \frac{
\rho(k) ( i k \wedge \varepsilon_{\lambda,k}) }{\sqrt{2|k|}} \left(
   a_{\lambda,k} e^{ik \cdot x}
   -
a_{\lambda,k}^* e^{-ik \cdot x} \right) dk \; .
$$
The
Hamiltonian is translation invariant and commutes with the generator
of translations, i.e., the operator of total momentum
$$
P_{\rm tot} = p + P_f  \; .
$$
Let $F$ be the Fourier transform in the electron variable $x$, i.e.,
on $L^2(\R^3)$,
$$
( F \psi)(\xi ) = \frac{1}{(2\pi)^{3/2}} \int_{\R^3} e^{-i \xi \cdot
x } \psi(x) dx \; .
$$
Set
$$
W = \exp(ix \cdot P_f) \; .
$$
Note $W P_{\rm tot} W^* = p $ so that in the new representation  $p$
is the total momentum. We compute
$$
W H W^* =  \left\{ \sigma \cdot ( p - P_f  +  e A ) \right\}^2 + H_f
\; ,
$$
where $A := A(0)$. Then the composition $U=FW$ yields the fiber
decomposition of the Hamiltonian and the Hilbert space
$$
U H U^* = \int_{\R^3}^{\oplus} H(\xi) d\xi \quad , \quad U : \HH \to L^2(\R^3; \C^2 ) \otimes \FF
\cong \int_{\R^3}^\oplus   \C^2  \otimes \FF  d\xi
$$
with
$$ H(\xi) = \left\{ \sigma \cdot ( \xi - P_f  +  e A )
\right\}^2 + H_f
 \;
$$
an operator on $\widetilde{\FF} := \C^2 \otimes \FF$. Note that
$H(\xi)$ can also be written as
$$
H(\xi) = (\xi - P_f + e A)^2 +  e \sigma \cdot B + H_f \; ,
$$
where $B := B(0)$. 
The explicit self-adjoint realization of $H(\xi)$ is given by the
following Lemma.
\begin{lemma} \label{lem:op1} The operator $H(\xi)$ is self-adjoint on
$D( P_f^2 + H_f ) = \{ \psi \in \widetilde{\FF} | (P_f^2 + H_f )
\psi \in \widetilde{\FF} \}$ and essentially self-adjoint on any
core of $P_f^2 + H_f$.
\end{lemma}
For a proof of Lemma \ref{lem:op1} see \cite{H00}. The operator
$H(\xi)$ is bounded from below and we write
$$
E(\xi) := \inf \sigma(H(\xi)) \; .
$$
\begin{proposition} \label{thm:lip1} The function $E(\cdot)$
is almost everywhere differentiable.
\end{proposition}
By spherical symmetry $E(\cdot) $ is invariant under rotations. We
want to point out that for small $e$ and $|\xi| < \frac{1}{6}$, it
has been shown that $E(\cdot)$ is twice differentiable with positive
Hessian \cite{C01,BCFS05}.
In \cite{F73,SPOHN04} it is shown that
for large $\xi$, $E(\xi) = |\xi| + O(1)$.  It seems probable that
for all $e$ and $\xi \neq 0$, $E(\cdot)$ is differentiable with
non-vanishing derivative.

\begin{theorem} \label{thm:main1} Let $e \neq 0$. If $E(\cdot)$
is differentiable at
 $\xi$ and has a
nonzero derivative, then $H(\xi)$ does not have a ground state
\end{theorem}

We want to relate this to  results obtained in \cite{C01,CF06},
where $A = A(0)$ in \eqref{eq:44} is replaced by an infrared
regularized $A_\sigma(0) = \sum_\lambda \int_{\sigma \leq |k|}
\rho(k) (2|k|)^{-1/2} ( a_{\lambda,k}^* \varepsilon_{\lambda,k} +
a_{\lambda,k} \varepsilon_{\lambda,k}) dk $. It is shown that if $e$
is small and $|\xi| < \frac{1}{6}$  then for any $\sigma
>0$, there exists a normalized ground state $\psi_\sigma(\xi)$. For
$\xi=0$, $\psi_\sigma(0)$ converges weakly as  $\sigma \to 0$ to a
nonzero vector. However for nonzero $\xi$, with $|\xi|<
\frac{1}{6}$, it was shown that $\psi_\sigma(\xi)$ converges weakly
to zero. We want to note that in principle this does not rule out
the possibility that there could suddenly appear a ground state in
Fock space at $\sigma=0$.

\section{The Electron: Proof of Results} \label{sec:proofelec}

\label{sec:elpro} First we give a well known proof of Lemma
\ref{thm:lip1}, see \cite{F73}.

\noindent {\it Proof of Proposition  \ref{thm:lip1}.} We set
$$
T(\xi) := H(\xi) - \xi^2  = - 2 \xi \cdot(P_f - e A) + (P_f - e A)^2
+ e \sigma \cdot B + H_f \; .
$$
Since for each $\psi \in D(P_f^2 + H_f) = D(H(\xi))$, the function
$\xi \mapsto (\psi, T(\xi)\psi)$ is linear, it follows that the
function
$$
\xi \mapsto t(\xi) := \inf \{   (\psi, T(\xi) \psi) | \psi \in
D(H(\xi)), \|\psi \| = 1 \}$$
 is concave. From concavity it follows that  $t(\cdot)$ is a.e.
 differentiable and hence also
the function $\xi \mapsto E(\xi)=\xi^2+t(\xi)$. \qed

\vspace{0.5cm}

\noindent For notational convenience we write
$$
v(\xi) = ( \xi - P_f + e A)  \; .
$$
Before we present the proof of Theorem \ref{thm:main1}, we need a
few Lemmas. For $E(\cdot)$ differentiable at $\xi$ and $\epsilon>0$,
we fix $\xi$ and consider the following subset of the unit sphere,
$$
S_\epsilon := \{ \omega \in S^2 |  \  \omega \cdot \nabla E(\xi)
\leq 1 -\epsilon \} \; .
$$
We denote normalized vectors by  $\widehat{k} = k /|k |$.

\begin{lemma} \label{lem:o(k)} Assume that $E(\cdot)$ is
differentiable at $\xi$.
For $\widehat{k}   \in S_\epsilon $ , we have
 $$
H(\xi - k) + |k| - E(\xi) \geq  \epsilon |k| + o(|k|)  \; .
$$
\end{lemma}

\begin{proof} Using that $E(\xi - k)$ is a lower bound for
$H(\xi - k)$ and the differentiability of $E(\cdot)$ at $\xi$,
 we have
\begin{eqnarray*}
H(\xi - k) + |k| - E(\xi) \geq E(\xi - k) - E(\xi) + |k| = - k \cdot
\nabla E(\xi) + |k| + o(|k|)  \geq \epsilon |k| + o(|k|) \; .
\end{eqnarray*}
\end{proof}
Let $P_0 = P_0(\xi)$ denote the orthogonal projection onto the
kernel of $H(\xi) - E(\xi)$. For $\varphi \in \widetilde{\FF}$, we
set
\begin{eqnarray} \label{eq:annihilation}
(a_{\underline{k}} \varphi )_n(\underline{k}_1, ... ,
\underline{k}_n) = (n+1)^{1/2} \varphi_{n+1}(\underline{k},
\underline{k}_1, ... , \underline{k}_n ) \; .
\end{eqnarray}
For $\lambda=1,2$, a.e. $k$, and all $n$, $
(a_{\underline{k}}\varphi)_n \in  S_n (\hh^{\otimes n} ) \otimes
\C^2$. The relation to $a(f)$ is outlined in the following Lemma.
\begin{lemma} \label{lem:new22}
Let $\Omega  \subset \R^3$ and $\varphi \in \widetilde{\FF}$ and
suppose the function $\underline{k} \mapsto a_{\underline{k}}
\varphi$ is in $L^2( \Z_2 \times \Omega; \widetilde{\FF})$. Then for
all $f \in \hh$, with $f$ vanishing outside of   $\Z_2 \times
\Omega$, and $\eta \in \widetilde{\FF}$
$$
( \eta , a(f) \varphi ) = \int \overline{f}(\underline{k}) ( \eta ,
a_{\underline{k}} \varphi ) d\underline{k} \; .
$$
\end{lemma}
\begin{proof}
We have
\begin{eqnarray*}
(\eta , a(f) \varphi ) && =  \sum_{n=0}^\infty \int \left( {\eta}_n(
\underline{k_1},  ... , \underline{k}_{n}), (n+1)^{1/2}
\overline{f}(\underline{k}) \varphi_{n+1}( \underline{k},
\underline{k}_1, ... , \underline{k}_{n+1})  \right) d \underline{k}
d\underline{k}_1 ...
d\underline{k}_n  \\
&& = \int  \overline{f}(\underline{k} ) \left( \sum_{n=0}^\infty
\int \left( {\eta}_n( \underline{k}_1, ... , \underline{k}_{n}),
(n+1)^{1/2} \varphi_{n+1}( \underline{k}, \underline{k}_1, ... ,
\underline{k}_{n+1}) \right)
d\underline{k}_1 ...  d\underline{k}_n \right) d\underline{k} \\
&& =  \int \overline{f}(\underline{k}) ( \eta , a_{\underline{k}}
\varphi ) d\underline{k} \; ,
\end{eqnarray*}
where the interchange of the order of integration and summation is
justified since
\begin{eqnarray*}
&& \sum_{n=0}^\infty \int \left|\left( {\eta}_n( \underline{k_1},
... , \underline{k}_{n}), (n+1)^{1/2} \overline{f}(\underline{k})
\varphi_{n+1}( \underline{k}, \underline{k}_1, ... ,
\underline{k}_{n+1}) \right) \right|  d \underline{k}
d\underline{k}_1 ... d\underline{k}_n   \\
&& \quad \quad  \leq  \| f \| \| \eta \| \left( \int_\Omega \|
a_{\underline{k}} \varphi \|^2 d \underline{k} \right)^{1/2} <
\infty \; .
\end{eqnarray*}
\end{proof}
\begin{lemma} \label{lem:new23} For each $\varphi \in D(H_f^{1/2})$,
the function $\underline{k}  \mapsto a_{\underline{k}} \varphi$ is
in $L^2_{\rm loc}(\Z_2 \times \R_{\times}^3  ; \widetilde{\FF} )$,
with $\R^3_\times = \R^3 \setminus \{ 0 \}$.
\end{lemma}

\begin{proof}
Since $\varphi \in D(H_f^{1/2})$, we conclude that
\begin{eqnarray*}
\lefteqn{\sum_{n=0}^\infty \int |k| | ( a_{\underline{k}}
\varphi)(\underline{k}_2, ..., \underline{k}_{n+1})
|^2 d\underline{k}_2 ...  d\underline{k}_{n+1} d\underline{k}} \\
&& = \sum_{n=0}^\infty \int \sum_{j=1}^{n+1} |k_j| |
\varphi_{n+1}(\underline{k}_1, \underline{k}_2, ... ,
\underline{k}_{n+1}) |^2  d\underline{k_1} d\underline{k}_2 ...
d\underline{k}_{n+1}  \\
&& = \| H_f^{1/2} \varphi \|^2 < \infty \; .
\end{eqnarray*}
This implies that the function $\underline{k} \mapsto \|
a_{\underline{k}} \varphi \|^2$ is integrable over any compact
subset of $\Z_2 \times \R_{\times}^3$.
\end{proof}
The next result uses the so called pull-through formula (see for
example \cite{F73}).
\begin{lemma} \label{lem:basicom1} Suppose $E(\cdot)$ is
differentiable at $\xi$ and and that $\psi$ is a  ground state of
$H(\xi)$.  Let $\epsilon > 0$. Then there exists a $\delta>0$ such
that for all $\eta \in \widetilde{ \FF} $,
\begin{equation} \label{eq:last888}
\left( \eta , a_{\underline{k}} \psi \right) =
 \frac{e \rho(k)}{\sqrt{2|k|} }
\left( H(\xi , k )^{-1} \eta, \left( - 2  \epsilon_{\underline{k}}
\cdot v(\xi) + i (  k \wedge  \epsilon_{\underline{k}})  \cdot
\sigma \right) \psi \right) \; ,
\end{equation}
for a.e. $k$, with $0< |k| < \delta $ and $\widehat{k} \in
S_\epsilon$, where $H(\xi,k) := H(\xi - k) + |k| - E(\xi)$.
\end{lemma}

\begin{proof} Let $f \in C_0^\infty(\R^3 \setminus \{ 0 \} )$ .
Let $\varphi \in \ran (P_{[0,\nu]}(N))$ be a state having less or
equal to $\nu$ photons, for some finite $\nu$, and assume each
$\varphi_n$ has compact support. By a calculation using the
canonical commutation relations, we find for real $f$,
\begin{eqnarray*}
\left( ( a_\lambda^*(f) H(\xi,k) - (H(\xi) - E(\xi)) a_\lambda^*(f)
) \varphi , \psi \right) = \left( ( A^*(f) + R_0^*(f)    + R_1^*(f)
) \varphi , \psi \right) \; ,
\end{eqnarray*}
with
\begin{eqnarray*}
R_0(f) &&:= \int  f(y) 2 ( y - k) \cdot  v(\xi)  a_{\lambda , y} \,
dy
 +
\int   f(y) ( k^2 - y^2) a_{\lambda,y} \, dy
 + \int  f(y) ( |k| - | y| ) a_{\lambda,y} \, dy \\
R_1(f) &&:=   \int  f(y)   \frac{ e \rho(y)}{\sqrt{2|y|}}
( k \cdot \varepsilon_{\lambda,y} )  \, dy \\
A(f) &&:= \int   f(y) \frac{e \rho(y)}{\sqrt{2|y|}} \left( - 2
\varepsilon_{\lambda,y} \cdot  v(\xi) + i (  y \wedge
\varepsilon_{\lambda,y}) \cdot  \sigma \right) \, dy  \; .
\end{eqnarray*}
Since $\psi \in D(H_f + P_f^2) \subset D(a_\lambda(f))$,
\begin{equation} \label{eq:proofgg}
( H(\xi,k) \varphi , a_\lambda(f) \psi ) = (\varphi , (A(f) + R_0(f)
+R_1(f)  ) \psi )
\end{equation}
Note that this holds for all $\varphi$ in an operator core for
$H(\xi,k)$.  For any $\epsilon
> 0$, there exists by Lemma \ref{lem:o(k)} a $\delta>0$ such that
for all $k$ with $0<|k|<\delta$ and $\widehat{k} \in S_\epsilon$,
$H(\xi,k)$ has a bounded inverse. This and equation
\eqref{eq:proofgg} imply that in fact for all such $k$ and all $\eta
\in \widetilde{\FF}$,
\begin{eqnarray} \label{eq:proofgg2}
 ( \eta , a_\lambda(f) \psi ) =
(H(\xi,k)^{-1} \eta , (A(f) + R_0(f) + R_1(f)  ) \psi ) \; .
\end{eqnarray}
Now fix $\eta \in \widetilde{\FF}$. For $k$, with $0<|k|<\delta$ and
$\widehat{k} \in S_\epsilon$, we choose a $\delta$--sequence
centered at $k$. Explicitly, we choose a nonnegative function $g \in
C_0^\infty(\R^3)$ with $\int g(y) dy =1$ and support in the unit
ball. We set $f_{k,m}(y) := m^3 g(m(y-k))$. By Lemmas
\ref{lem:new22} and \ref{lem:new23} it follows  that the left hand
side of \eqref{eq:proofgg2} yields, $\lim_{m\to \infty} (\eta ,
a_\lambda(f_{m,k}) \psi ) = (\eta, a_{\lambda,k} \psi)$ a.e. $k$.
The term $(H(\xi,k)^{-1}\eta, A(f_{m,k}) \psi)$ converges to the
right hand side of \eqref{eq:last888}.
 Below we  will show that the terms $(H(\xi,k)^{-1} \eta,
R_0(f_{m,k}) \psi)$ and $(H(\xi,k)^{-1} \eta, R_1(f_{m,k}) \psi)$
vanish as $m$ tends to infinity for a.e. $k$.
 The expression containing $R_1$  vanishes since
$k \cdot \epsilon_{\lambda,k} = 0$. To show that the expression
involving $R_0$ vanishes we will only consider one term. The other
terms will follow similarly. We set $\phi_l := v_l(\xi)
H(k,\xi)^{-1} \eta $ and estimate
\begin{eqnarray*}
\lefteqn{ R_{0,1}(f_{m,k}) := \left|\left( H(k,\xi)^{-1} \eta , \int
f_{k,m}( y) 2 ( y - k)
\cdot  v(\xi) a_{\lambda,y} \psi \, dy \right) \right| }  \\
&&\leq \sum_{l=1}^3  \left|\left(  v_l(\xi) H(k,\xi)^{-1} \eta ,
\int  f_{k,m}(y) 2 ( y - k)_l  a_{\lambda,y} \psi \,  dy \right) \right|  \\
&& \leq \sum_{l=1}^3 \sum_{n=0}^\infty \int  \left|
\left(({\phi}_l)_n(\underline{k}_1, ... , \underline{k}_n), \int
f_{k,m} (y) 2 ( y - k)_l (n+1)^{1/2} \psi_{n+1}(\lambda, y,
\underline{k}_1, ... , \underline{k}_n )  dy  \right) \right|
d\underline{k}_1 ...
d\underline{k}_n\\
&& \leq   \| \phi \| \int   f_{k,m} (y) 2 | y - k| h_\lambda(y) dy
\; ,
\end{eqnarray*} where
$$
h_\lambda(y) = \left( \sum_{n=0}^\infty  \int  (n+1) \left|
\psi_{n+1}(\lambda, y, \underline{k}_1, ... , \underline{k}_n )
\right|^2    d\underline{k}_1 ... d\underline{k}_n  \right)^{1/2} \;
$$
and $$ \| \phi \|^2  = \sum_{l=1}^3 \| \phi_l \|^2 \; .
$$
 Since $\psi$ is in $D(H_f^{1/2})$,
$$
\int \left| |y|^{1/2} h_\lambda(y) \right|^2 dy \leq
\sum_{n=0}^\infty   \sum_{\mu=1,2}  \int (n+1) |y| \left|
\psi_{n+1}(\mu, y, \underline{k}_1, ... , \underline{k}_n )
\right|^2 dy d\underline{k}_1 ... d\underline{k}_n  = ( \psi , H_f
\psi ) \; .
$$
Thus $h_\lambda \in L^1_{\rm loc}(\R^3 \setminus \{ 0 \} )$.
Therefore a.e. point is a Lebesgue point of $h_\lambda$. At such
points $k$,
$$ \int f_{k,m}(y) h_\lambda(y) dy \to h_\lambda(k) \; ,$$
by Lebesgue's differentiation theorem, see for example \cite{SW71}
Theorem 1.25. Thus $R_{0,1}(f_{k,m})$ tends to zero  as
 $m \to \infty$, a.e. $k$.
\end{proof}

\begin{lemma} \label{lem:nablae1} If $E(\cdot )$ is differentiable at $\xi$, then
$$
P_0 2 v(\xi) P_0 = \nabla E(\xi) P_0 \; .
$$
\end{lemma}

\begin{proof}
Suppose $\psi \in \ran P_0$, with $\|\psi \|=1$, then
\begin{eqnarray*}
E(\xi + k) - E(\xi) \leq ( \psi, (H(\xi + k) - H(\xi) ) \psi ) =  2
k \cdot ( \psi , v(\xi) \psi ) + |k|^2 \; .
\end{eqnarray*}
This implies
$$
k  \cdot \nabla E(\xi) \leq  2 k  \cdot (\psi , v(\xi) \psi ) +
o(|k|) \quad , \quad |k| \to 0 \; .
$$
Since $k$ can have any direction we conclude that
$$
\nabla E(\xi) = 2 ( \psi, v(\xi) \psi ) \; .
$$
Since this holds for any $\psi \in \ran P_0$ the claim follows by
polarization.
\end{proof}

We set
$$
Q(k) = |k| ( H(\xi - k) + |k| - E(\xi))^{-1} \; ,
$$
whenever this exists. And for $|k| > 0$, we set
$$
Q_0(k) = |k| ( H(\xi) + |k| - E(\xi) )^{-1} \; .
$$
By the spectral theorem
$$
P_0 = P_0(\xi) = s-\lim_{|k| \to 0} Q_0(k) \; .
$$

\begin{lemma} \label{lem:weak} Let $E(\cdot)$ be differentiable at $\xi$. Given
$\epsilon > 0$, then
$$
w- \lim_{\widehat{k} \in S_\epsilon, |k| \to 0} \left( Q(k) - ( 1 -
\widehat{k} \cdot \nabla E(\xi) )^{-1} P_0 \right) = 0 \; .
$$
\end{lemma}

\begin{proof}
Fix $\xi$

\vspace{0.5cm}

\noindent \underline{Step 1:} $v(\xi) Q_0(k)$ is uniformly bounded
for small $|k|$.

\vspace{0.5cm} Since $B$ is $H_f^{1/2}$ operator bounded, we see
that there exists a finite constant $C_0$ such that
\begin{eqnarray}
\label{ineq:55}
 v(\xi)^2 \leq   H(\xi) + C_0 \leq  ( H(\xi) +  |k| -
E(\xi) ) + ( E(\xi) + C_0 ) \; .
\end{eqnarray}
On the other hand
\begin{eqnarray*}
 \sum_{l=1}^3 ( v(\xi) Q_0(k))_l^* ( v(\xi) Q_0(k) )_l
= \frac{|k|}{H(\xi) - E(\xi) + |k| } v(\xi)^2 \frac{|k|}{H(\xi) -
E(\xi) + |k| } \; .
\end{eqnarray*}
By inequality \eqref{ineq:55} we see that the right hand side is
uniformly bounded for small $|k|$. This shows Step 1.

\vspace{0.5cm}

\noindent \underline{Step 2:} We have $s-\lim_{|k| \to 0} v(\xi)
Q_0(k) = v(\xi) P_0 $.

\vspace{0.5cm} \noindent By the resolvent identity
\begin{eqnarray}
\lefteqn{ v(\xi) \frac{|k|}{H(\xi) - E(\xi) + |k|} } \nonumber \\
&& = v(\xi)  \frac{|k|}{H(\xi) - E(\xi) + |k|  + 1} - v(\xi)
\frac{1}{H(\xi) - E(\xi) + |k| + 1 } \frac{|k|}{H(\xi) - E(\xi) +
|k|} \; .  \label{step2:conv1}
\end{eqnarray}
Again using the resolvent identity and an argument similar to the
one in Step 1,
\begin{eqnarray*}
\lefteqn{ \left\| v(\xi) \frac{1}{H(\xi) - E(\xi) + |k| + 1} -
v(\xi) \frac{1}{H(\xi) - E(\xi) + 1 } \right\| } \\
&& =\left\|  v(\xi) \frac{1}{H(\xi) - E(\xi)  + 1}  |k |
\frac{1}{H(\xi) - E(\xi) + |k| + 1 }   \right\| \\
&&  \stackrel{ |k| \to 0 }{\longrightarrow} 0 \; .
\end{eqnarray*}
This implies that the first term on the right hand side in
\eqref{step2:conv1} converges in norm to zero and the second term
converges strongly to $v(\xi)P_0$.

 \vspace{0.5cm}

\noindent \underline{Step 3:}  Uniformly for $\widehat{k} \in
S_\epsilon$,
$$P_0 Q(k) P_0 -  \left( P_0 2 \widehat{k} \cdot v(\xi) P_0 \right)
(P_0 Q(k) P_0 ) \stackrel{w}{\longrightarrow} P_0 \;  \ \  {\rm and}
\ \ \ Q(k) - P_0Q(k) P_0 \stackrel{w}{\longrightarrow} 0 \; .
$$

\vspace{0.5cm} \noindent Using the second resolvent identity twice
we obtain for small $|k|$ and $\widehat{k} \in S_\epsilon$,
\begin{eqnarray}
Q(k) &=& Q_0(k)  + Q_0(k) ( 2  \widehat{k}\cdot  v(\xi) - |k| ) Q(k)
\label{solventone1} \\
&=& Q_0(k) + Q_0(k)( 2 \widehat{k} \cdot v(\xi) - |k| ) Q_0(k)  \nonumber \\
&& + Q_0(k) ( 2 \widehat{k}\cdot v(\xi) - |k| ) Q(k )  ( 2
\widehat{k} \cdot  v(\xi) - |k| ) Q_0(k)   \label{resolventtwo1}
\end{eqnarray}
Now using \eqref{resolventtwo1} and the results of Step 1 and Step
2, we find
$$
Q(k) ( 1 - P_0 ) \stackrel{w}{\longrightarrow} 0 \quad , \quad ( 1 -
P_0 ) Q(k)
 \stackrel{w}{\longrightarrow} 0  \; ,
$$
where the limit is uniform for $\widehat{k} \in  S_\epsilon$. It
follows that
$$
Q(k) - P_0Q(k) P_0 \stackrel{w}{\longrightarrow} 0 \; ,
$$
uniformly for $\widehat{k} \in S_\epsilon$. Now this and
\eqref{solventone1} show Step~3.

 \vspace{0.5cm}

The claim of the Lemma is now an immediate consequence of Lemma
\ref{lem:nablae1}  and Step~3.
\end{proof}

Now we are ready to prove Theorem \ref{thm:main1}.

\vspace{0.5cm}

\noindent {\it Proof of Theorem \ref{thm:main1}.} Suppose $H(\xi)$
has a ground state $\psi$ with $\| \psi \| = 1$. We want to show
this leads to a contradiction. We  choose an $\eta \in
D((N+1)^{1/2})$ such that $( \eta ,\psi) \neq 0$. Choose $\epsilon$
with $0< \epsilon < 1$ and $\delta >0$ sufficiently small. Then  by
Lemma \ref{lem:basicom1}  for a.e. $k$ with  $\widehat{k} \in
S_\epsilon$ and $|k| < \delta$,
\begin{eqnarray*}
 (\eta , a_{\lambda,k}  \psi ) && =   \left( \eta , H(\xi,k)^{-1}
({2|k|})^{-1/2} e \rho(k) \left( - 2 \varepsilon_{\lambda,k} \cdot
v(\xi) +
 i ( k \wedge \varepsilon_{\lambda, k}  ) \cdot \sigma \right) \psi
\right) \\
&& = \frac{ e \rho(k) } {\sqrt{2}|k|^{3/2}}  \left( \eta , Q(k)
\left( -  2 \varepsilon_{\lambda,k} \cdot v(\xi) + i ( k \wedge
\varepsilon_{\lambda, k}  ) \cdot \sigma \right) \psi \right) \; .
\end{eqnarray*}
Now uniformly for $\widehat{k} \in S_\epsilon$,
\begin{eqnarray*}
\left( \eta , Q(k) \left( - 2  \varepsilon_{\lambda,k} \cdot v(\xi)
+ i ( k \wedge \varepsilon_{\lambda, k} ) \cdot \sigma \right) \psi
\right) \stackrel{|k| \to 0}{\longrightarrow} && -
 (1 - \widehat{k} \cdot \nabla E )^{-1}  \varepsilon_{\lambda,k} \cdot (P_0 \eta,
2 v(\xi) \psi ) \\
&&  = -  ( \varepsilon_{\lambda,k} \cdot \nabla E ) (1 - \widehat{k}
\cdot  \nabla E )^{-1}
 (\eta, \psi ) \; ,
\end{eqnarray*}
where in the last step we used  Lemma \ref{lem:nablae1}.
 We introduce the set
$$
K := \{ \omega \in S^2 | - \sfrac{1}{2} | \nabla E | \leq \omega
\cdot \nabla E \leq 0 \} \subset S_\epsilon \; .
$$
Then there exists a positive constant $c_0$ such that for all
$\widehat{k} \in K$,
$$
\sum_{\lambda=1,2} | ( \varepsilon_{\lambda, k} \cdot \nabla E ) |^2
\geq c_0
> 0  \; .
$$
By the above, there exists a nonzero $\delta_2$ such that for a.e.
$k$ with $|k|<\delta_2$ and $\widehat{k} \in K$,
\begin{eqnarray*}
\sum_{\lambda=1,2} | ( \eta , a_{\lambda , k} \psi ) |^2 \geq
\frac{1}{2} \frac{ |e \rho(k)|^2}{{2}|k|^{3}} ( 1 - \widehat{k}
\cdot \nabla E )^{-2} \left| ( \eta , \psi )\right|^2 c_0 \; .
\end{eqnarray*}
Therefore, there exists a $c_1 > 0$ such that for a.e. small $k$
with $\widehat{k} \in K$, we have
$$
\frac{| e \rho(k) c_1 |^2}{|k|^{3}} \leq \sum_{\lambda=1,2}| ( \eta
,a_{\lambda,k} \psi ) |^2 \leq \| ( 1 + N)^{1/2} \eta \|^2 \left(
\sum_{\lambda=1,2} \sum_{n=0}^\infty  \int | \psi_{n+1}(\lambda, k,
\underline{k}_1,  ... , \underline{k}_n )|^2 d\underline{k}_1 ...
d\underline{k}_n \right)  \; .
$$
Integrating over the set of all $k$ with $\widehat{k} \in K$ and
$|k| \leq \delta_2$,  we see this is inconsistent with $\psi$ being
in $\widetilde{\FF}$. Thus $H(\xi)$ does not have a ground state.
 \qed

\section{Positive Ion: Model and Statement of Results}

\label{sec:iondef}

We consider an ion consisting of a spinless nucleus of mass $m_0$
and charge $Ze$ and $N$ spin $1/2$ electrons having charge $-e$ and
mass $1$. The energy of this system is described by  the operator
$$
H = \frac{1}{2m_0} \left( p_0 - Z e A(x_0) \right)^2 + \sum_{j=1}^N
\frac{1}{2}\{ \sigma_j \cdot (p_j + e A(x_j) ) \}^2 + H_f +  V(x_0,
... , x_N) \; ,
$$
acting on the Hilbert space
$$
\HH = L^2(\R^3 ) \otimes  \left( \bigwedge^N_{j=1} L^2(\R^3; \C^2 )
\right)  \otimes \FF \; ,
$$
where $p_0= - i \nabla_{0}$ acts on the first factor  and $p_j = - i
\nabla_j$ and $\sigma_j$, the three-vector of Pauli matrices, act on
the $j$-th factor of the antisymmetric tensor product. We take the
spin of the nucleus to be zero only to simplify notation. We will
make the following assumptions about the potential $V$:
\begin{eqnarray*}
V(x_0, ... , x_N) =
\sum_{0 \leq i < j \leq N} V_{ij}(x_i - x_j ) \; .
\end{eqnarray*}
Each $V_{ij}$ is infinitesimally bounded with respect to the
Laplacian in three dimensions,
which we denote by $-\Delta$, i.e.,
there exists for any $a > 0$ a  finite constant $b$ such that for
 all $f$ in the domain
of $-\Delta$,
$$
\| V_{ij} f \| \leq a  \|-\Delta f \| + b \| f \| \ \; .
$$
The Hamiltonian is translation
invariant and therefore commutes with the generator of translations,
i.e., the operator of total momentum
$$
P_{\rm tot} = \sum_{j=0}^N  p_j + P_f \; .
$$
Let $F$ be the Fourier transform
in the variable $x_0$, i.e., on $L^2(\R^3)$,
$$
( F \psi)(\xi ) = \frac{1}{(2\pi)^{3/2}} \int_{\R^3} e^{-i \xi \cdot
x_0 } \psi(x_0) dx_0 \; .
$$
Let
$$
W = \exp(ix_0 \cdot (P_f  + \sum_{j=1}^N p_j ) )  \; .
$$
Note that $W P_{\rm tot} W^* = p_0$ so that in a new representation,
$p_0$ is the total momentum. Then the composition $U=FW$ yields the
decomposition of the Hamiltonian
$$
U H U^* = \int_{\R^3}^{\oplus} H(\xi) d\xi \; ,
$$
with
$$
H(\xi) = \frac{1}{2m_0}( \xi - \sum_{j=1}^N p_j - P_f  - Z e A(0)
)^2 +  \frac{1}{2} \sum_{j=1}^N \left\{ \sigma_j \cdot ( p_j + e
A(x_j) ) \right\}^2 + H_f + \widetilde{V} \;
$$
acting on $\left( \bigwedge^N L^2(\R^3; \C^2) \right) \otimes \FF $
and where we have set $\widetilde{V} = V\vert_{x_0=0}$. Let us cite
the following Theorem \cite{LMS06,H02}.
\begin{theorem} \label{thm:domain2} The operator $H(\xi)$ is self-adjoint on
$$ \bigcap_{j=1}^N D( p_j^2 ) \cap D(P_f^2 + H_f )$$
 and essentially
self-adjoint on any core of $\sum_{j=1}^N p_j^2  + P_f^2 + H_f$.
\end{theorem}
It is easy to show that for every $\xi$ the operator $H(\xi)$ is
bounded below. Let $E(\xi) = \inf \sigma( H(\xi))$ be the infimum of
the spectrum. By a simple argument as in the proof of Proposition
\ref{thm:lip1} we see that $E(\cdot)$ is almost everywhere
differentiable. The following theorem is the main result. Its proof
is given in the next section.
\begin{theorem} \label{thm:main22} Suppose $N \neq Z$ and $e\neq 0$.
If $E( \cdot )$ is differentiable at $\xi$ with non-vanishing
derivative then $H(\xi)$ does not have a ground state.
\end{theorem}

\section{Positive Ion: Proof of Result}

\label{sec:ionpro}

First we show the following lemma.

\begin{lemma}   \label{lem:formbound} $|\widetilde{V}|$ is
infinitesimally form bounded with respect to  $H(\xi)$.
\end{lemma}

\begin{proof} By Theorem \ref{thm:domain2}, we know that
$H(\xi)$ is self-adjoint on the domain of $P_f^2 + \sum_{j=1}^Np_j^2
+ H_f$. Therefore there exist finite constants $c_1$ and $c_2$ such
that
$$
P_f^2 + \sum_{j=1}^Np_j^2 + H_f \leq c_1 H(\xi) + c_2  \; .
$$
By assumption $\widetilde{V}$ is infinitesimally small with respect
to $\sum_{j=1}^N p_j^2$. Therefore, $|\widetilde{V}|$ is
infinitesimally form bounded with respect to  $\sum_{j=1}^N p_j^2$.
Hence  for any $a>0$ there exists a finite $b$ such that
\begin{eqnarray*}
|\widetilde{V}| \leq  a  \sum_{j=1}^N p_j^2 +  b \leq a \left( P_f^2
+ \sum_{j=1}^N p_j^2 + H_f \right) + b  \leq a c_1 H(\xi) + a c_2 +
b \; .
\end{eqnarray*}
\end{proof}
We will prove Theorem \ref{thm:main22} using a sequence of Lemmas.
For notational convenience we set
$$
v(\xi) =  \xi - \sum_{j=1}^N p_j - P_f - Z e A(0) \; .
$$
Recall the definitions $S_\epsilon := \{ \omega \in S^2 | \ \omega
\cdot \nabla E(\xi) \leq 1 - \epsilon \}$ and  $\widehat{k} :=
k/|k|$, which are the same as in Section \ref{sec:proofelec}.
\begin{lemma}\label{lem:E(k)2} Assume that $E(\cdot)$ is differentiable at $\xi$. Given $
\epsilon > 0$, then for $\widehat{k}  \in S_\epsilon$,  we have
 $$
H(\xi - k) + |k| - E(\xi) \geq  \epsilon |k| + o(|k|)  \; .
$$
\end{lemma}
The proof of  Lemma \ref{lem:E(k)2} is the same as the proof of
Lemma \ref{lem:o(k)}.
\begin{lemma} \label{lem:new222} Let $\HH_0$ be any Hilbert space.
Let $\Omega  \subset \R^3$ and $\varphi \in  \HH_0 \otimes \FF$, and
suppose the function $\underline{k} \mapsto a_{\underline{k}}
\varphi$ is in
 $L^2(\Z_2 \times \Omega; \HH_0 \otimes \FF)$. Then for all $f \in \hh$, with $f$ vanishing outside of  $\Z_2 \times \Omega$,
 and $\eta \in \HH_0 \otimes \FF$
$$
( \eta , a(f) \varphi ) = \int \overline{f}(\underline{k}) ( \eta ,
a_{\underline{k}} \varphi ) d\underline{k} \; .
$$
\end{lemma}
The proof of this Lemma is analogous to the proof of  Lemma
\ref{lem:new22}. We merely have to replace the inner product of
$\mathbb{C}^2$ by the inner product of $\HH_0$. Likewise, one
generalizes the proof of Lemma \ref{lem:new23} to prove  the next
lemma. Anticipating our application we set henceforth $\HH_0:=\left(
\bigwedge^N L^2(\R^3; \C^2) \right) $.
\begin{lemma} \label{lem:new233} Let $\varphi \in D(H_f^{1/2})$.
Then the  function $\underline{k}  \mapsto  a_{\underline{k}}
\varphi$ is in $L^2_{\rm loc}(\Z_2 \times \R_\times^3  ; \HH_0
\otimes \FF )$, with $\R^3_\times = \R^3 \setminus \{ 0 \}$.
\end{lemma}

\begin{lemma}
\label{lem:basicom} Suppose $E(\cdot)$ is differentiable at $\xi$
and  that $\psi$ is a  ground state of $H(\xi)$. Let $\epsilon > 0$.
Then there exists a $\delta > 0$ such that for all $\eta \in \HH_0
\otimes \FF$,
\begin{eqnarray}
\lefteqn{ \left(  \eta  , a_{\lambda,k} \psi   \right) }
\label{eq:lem:basicom}
\\ &&=      \frac{e \rho(k)}{\sqrt{2|k|}}
 \left( H(\xi,k)^{-1} \eta  ,  \left(  \frac{Z}{m_0} v(\xi)
-  \sum_{j=1}^N  e^{-ik \cdot x_j} (  \frac{1}{2} i  k \wedge
\sigma_j + p_j + e A(x_j) ) \right) \cdot \varepsilon_{\lambda,k}
\psi \right) \; , \nonumber
\end{eqnarray}
for a.e. $k$, with  $0 < |k| < \delta$ and $\widehat{k} \in
S_\epsilon$, where $H(\xi,k) := H(\xi - k) + |k| - E(\xi)$.
\end{lemma}
\begin{proof} Let $f \in C_0^\infty(\R^3 \setminus \{ 0 \} )$.
Let $\varphi \in \ran (P_{[0,\nu]}(N))$ be a state having less or
equal to $\nu$ photons, for some finite  $\nu$, and assume
$\varphi_n$ is smooth and has compact support. Then a
straightforward calculation using the canonical commutation
relations, yields for $f$ real,
\begin{eqnarray*}
\left( ( a^*_\lambda(f) H(\xi,k) - (H(\xi) - E(\xi) )a^*_\lambda(f)
) \varphi , \psi \right) = \left( ( A^*(f) + R_0^*(f) + R^*_1(f) )
\varphi , \psi \right) \; ,
\end{eqnarray*}
with
\begin{eqnarray*}
R_0(f) &&:= \int ( |k| - |y|)  f(y) a_{\lambda,y} \, dy
 + m_0^{-1}  \int f(y) (y-k) \cdot v(\xi) a_{\lambda,y} \, dy \\
 &&
+  (2 m_0)^{-1} \int  f(y) ( k^2 - y^2 )   a_{\lambda,y}  \, dy \\
R_1(f) &&:=  -    \frac{Z}{2m_0} e  \int  \frac{
\rho(y)}{\sqrt{2|y|}} f(y)  k  \cdot \varepsilon_{\lambda,y} \, dy
\\
A(f) &&:=   -  \sum_{j=1}^N  e \int  \frac{ \rho(y)}{\sqrt{2|y|}}
e^{-i y \cdot x_j } f(y)  \varepsilon_{\lambda,y}  \cdot (p_j + e
A(x_j)) \, dy \\
&&
 +   \frac{Z}{m_0} e \int  \frac{ \rho(y)}{\sqrt{2|y|}}   f(y)
  \varepsilon_{\lambda,y} \cdot v(\xi) \, dy \\
 &&  + \frac{1}{2} \sum_{j=1}^N e \int   \frac{ \rho(y)}{\sqrt{2|y|}}
 e^{- i y \cdot x_j} f(y)  ( i k \wedge \varepsilon_{\lambda,y} )
 \cdot  \sigma_j  \, dy
 \; .
\end{eqnarray*}
Since $\psi \in \bigcap_{j=1}^N D(p_j^2) \cap D(P_f^2 + H_f)
\subset D(a_\lambda(f))$,
$$( H(\xi,k) \varphi , a_\lambda(f) \psi )
= (\varphi , (A(f) + R_0(f) + R_1(f) ) \psi ) \;.$$
Note that this holds for all $\varphi$ in an operator core for
$H(\xi,k)$. For $\epsilon > 0$, there exists by Lemma
\ref{lem:E(k)2} a $\delta
> 0$ such that for all $k$ with $0<|k|<\delta$ and $\widehat{k} \in
S_\epsilon$, $H(\xi,k)$ has a bounded inverse. Thus we conclude by
density that for all such $k$ and all $\eta \in \HH_0 \otimes \FF$,
\begin{equation} \label{eq:lhs}
( \eta , a_\lambda(f) \psi ) = (H(\xi,k)^{-1}  \eta  , (A(f) +
R_0(f) + R_1(f) ) \psi ) \; .
\end{equation}
Now fix $\eta \in \HH_0 \otimes \FF$. For  $k$, with $0<|k|<\delta$
and $\widehat{k} \in S_\epsilon$, we choose a $\delta$-sequence,
$f_{m,k}$, centered at $k$ as in the proof of Lemma
\ref{lem:basicom1}. We insert $f_{m,k}$ for $f$ in equation
\eqref{eq:lhs}. As $m \to \infty$, it follows by Lemmas
\ref{lem:new222} and \ref{lem:new233} that the left hand side of
\eqref{eq:lhs} converges to the left hand side of
\eqref{eq:lem:basicom} for a.e. $k$. In the same limit the term
involving $A$ converges to the right hand side of
\eqref{eq:lem:basicom}. As demonstrated in the proof of Lemma
\ref{lem:basicom1} the terms involving $R_0$ and $R_1$ vanish as $m$
tends to infinity for a.e. $k$. This implies the assertion of the
Lemma.
\end{proof}
The next lemma would follow easily from the formal commutation
relation
$$
[H(\xi), i x_j] = -  \frac{1}{m_0} v(\xi) +    p_j + e A(x_j) \;
$$
if we ignored domain considerations.

\begin{lemma} \label{lem:virial} Let $P_0$ be the projection onto the kernel
of $H(\xi) - E(\xi)$. Then for all  $j$ with $1 \leq j \leq N$,
$$
P_0   \frac{1}{m_0} v(\xi) P_0 = P_0   ( p_j + e A(x_j) )P_0 \; .
$$
\end{lemma}

\begin{proof} Fix a $j \in \{ 1,2, ... , N\}$.
Let $\chi \in C^\infty(\R_+ ;[0,1])$ with $\chi
\restricted{[0,1]}=1$ and $\chi \restricted {[2,\infty)} = 0$. We
set $\chi_n(x_j) = \chi(|x_j|/n)$. Let $\psi \in \ran P_0$, then for
all $n$
\begin{eqnarray*}
0 && = \langle \psi , H(\xi) i \chi_n (x_j) x_j \psi \rangle -
\langle \psi , i \chi_n(x_j) x_j H(\xi) \psi \rangle \\
 && = \langle
\chi_n(x_j) \psi , (- \frac{1}{m_0} v(\xi) + p_j + e A(x_j) ) \psi
\rangle \\
&&
 +  {\rm Re} \langle \psi , \frac{1}{n} (\nabla \chi)(|x_j|/n) x_j \cdot (-
\frac{1}{ m_0} v(\xi) + p_j + e A(x_j) ) \psi \rangle \\
&& - i \left( \psi , \left( \frac{1}{2 m_0} + \frac{1}{2} \right)
\frac{1}{n} (\nabla
\chi)(|x_j|/n) \psi \right) \\
  &&\stackrel{n\to \infty}{\longrightarrow} \langle \psi ,  (-
\frac{1}{m_0} v(\xi) +    p_j + e A(x_j)  ) \psi \rangle \; .
\end{eqnarray*}
The limit as $n$ tends to infinity follows from dominated
convergence.
 By polarization this yields the
claim.
\end{proof}
The proof of the next lemma is the same as the proof of Lemma
\ref{lem:nablae1}.
\begin{lemma} \label{lem:nablae} Let $P_0$ be the projection onto the
the kernel of $H(\xi) - E(\xi)$. If $E(\cdot )$ is differentiable at
$\xi$, then
$$
P_0 \frac{1}{m_0} v(\xi) P_0 = \nabla E(\xi) P_0 \; .
$$
\end{lemma}

We set
$$
Q(k) = |k| ( H(\xi - k) + |k| - E(\xi))^{-1} \; ,
$$
whenever this exists. And for $|k| > 0$, we set
$$
Q_0(k) = |k| ( H(\xi) + |k| - E(\xi) )^{-1} \; .
$$
Let $P_0$ be the orthogonal projection onto the kernel of $H(\xi) -
E(\xi)$. By the spectral theorem
$$
P_0 = P_0(\xi) = s-\lim_{|k| \to 0} Q_0(k) \; .
$$

\begin{lemma} \label{lem:weak2} Let $E(\cdot)$ be differentiable at $\xi$. Given
$\epsilon > 0$. Then for $\widehat{k} = k /|k|$,
$$
w- \lim_{\widehat{k} \in S_\epsilon, |k| \to 0} \left( Q(k) - ( 1 -
\widehat{k} \cdot \nabla E(\xi) )^{-1} P_0 \right) = 0 \; .
$$
\end{lemma}
The proof follows the steps of Lemma \ref{lem:weak}, where  Step 1
uses Lemma \ref{lem:formbound}. We now present the proof of Theorem
\ref{thm:main22}.

 \vspace{0.5cm}

\noindent {\it Proof of Theorem \ref{thm:main22}.} Suppose $H(\xi)$
has a ground state $\psi$ with $\| \psi \| = 1$. We want to show
that this leads to a contradiction. Choose $\epsilon$ with $0<
\epsilon < 1$, and choose $\eta \in D((N+1)^{1/2})$ with
$(\eta,\psi) \neq 0$. By Lemma \ref{lem:basicom} there exists a
$\delta
>0$ such that for a.e. $k$, with $0<|k|<\delta$ and $\widehat{k} \in
S_\epsilon$,
\begin{eqnarray*}
( \eta , a_{\lambda,k} \psi ) &&= \frac{e
\rho(k)}{\sqrt{2}|k|^{3/2}} \left[ \left( \eta , Q(k) \frac{1}{2}
\sum_{j=1}^N e^{- i k \cdot x_j }( i k \wedge
\varepsilon_{\lambda,k} ) \cdot \sigma_j \psi
\right) \right. \\
&& + \left. \varepsilon_{\lambda , k } \cdot \left\{ \left( \eta ,
Q(k)\frac{Z}{m_0} v(\xi) \psi \right)  +    \left( \eta , Q(k)
\sum_{j=1}^N ( - e^{-ik \cdot x_j})(p_j + e A(x_j) ) \psi \right)
\right\} \right] \; .
\end{eqnarray*}
Since $Q(k)$ is uniformly bounded on $S_\epsilon$  for small $|k|$,
$$
\left( \eta , Q(k)   \sum_{j=1}^N   e^{-ik \cdot x_j} i (k \wedge
\varepsilon_{\lambda,k} ) \cdot \sigma_j \psi \right) \stackrel{|k|
\to 0}{\longrightarrow} 0 \; \; ,
$$
uniformly for  $k \in S_\epsilon$.
 Using Lemma \ref{lem:weak}, we find uniformly for $\widehat{k} \in
 S_\epsilon$
as $|k| \to 0$,
\begin{eqnarray*}
 \left( \eta ,
Q(k)\frac{Z}{m_0} v(\xi) \psi \right) {\longrightarrow}&& ( 1 -
\widehat{k} \cdot \nabla
E )^{-1} \left( P_0 \eta , \frac{Z}{m_0} v(\xi) \psi \right) \; \\
 && =  Z ( \nabla E ) (1 - \widehat{k} \cdot \nabla
E)^{-1} (\eta, \psi ) \; ,
\end{eqnarray*}
where we used Lemma \ref{lem:nablae}. Again by Lemma \ref{lem:weak}
and using that $e^{-ik \cdot x_j}$ converges in the strong operator
topology to $1$, we find uniformly for $k \in S_\epsilon$ as $|k|
\to 0$,
\begin{eqnarray*}
 \left( \eta , Q(k)
\sum_{j=1}^N ( - e^{-ik \cdot x_j})(p_j + e A(x_j) ) \psi \right)
 {\longrightarrow}   && ( 1 -  \widehat{k} \cdot \nabla E )^{-1}
\left( P_0
\eta , - \sum_{j=1}^N (p_j + e A(x_j))  \psi \right) \\
&&= ( 1 - \widehat{k} \cdot \nabla E )^{-1} \left( P_0 \eta ,
\frac{-N}{m_0} v(\xi)  \psi \right) \\
&& =  - {N}  (\nabla E) (  1 - \widehat{k} \cdot  \nabla E )^{-1}
(\eta ,\psi)
\end{eqnarray*}
where in the second line we used Lemma \ref{lem:virial} and in the
last again Lemma \ref{lem:nablae}.
 We introduce the set
$$
K := \{ \omega \in S^2 | - \sfrac{1}{2} | \nabla E | \leq \omega
\cdot \nabla E \leq 0 \} \subset S_\epsilon \; .
$$
Then, since by assumption $\nabla E \neq 0$, there exists a positive
constant $c_0$ such that for all   $\widehat{k} \in K$,
$$
\sum_{\lambda=1,2} | \varepsilon_{\lambda, k} \cdot  \nabla E  |^2
\geq c_0
> 0  \; .
$$
Collecting the above estimates we conclude that for small $|k|$
uniformly for  $\widehat{k} \in K$,
$$
\sum_{\lambda=1,2}| ( \eta , a_{\lambda , k} \psi ) |^2 \geq
\frac{1}{2} \frac{|e \rho(k)|^2}{2|k|^{3}} ( 1 - \widehat{k} \cdot
\nabla E )^{-2}  | Z - N |^2 | (\eta , \psi ) |^2 c_0 \; .
$$
By this and $N \neq Z$, there exists a $c_1 > 0$ such that for all
small $k$ with $\widehat{k} \in K$, we find
\begin{eqnarray*}
 \frac{|\rho(k) c_1 |^2}{|k|^{3}}  && \leq
\sum_{\lambda=1,2}| ( \eta ,a_{k,\lambda} \psi ) |^2 \\
&& \leq \| ( 1 + N)^{1/2} \eta \|^2 \left( \sum_{\lambda=1,2}
\sum_{n=0}^\infty \int \| \psi_{n+1}(\lambda,k, \underline{k}_1,
... , \underline{k}_n  ) \|^2 d\underline{k}_1 ... d\underline{k}_n
\right) \; ,
\end{eqnarray*}
where in the last inequality we used
 Cauchy-Schwarz.
 This is inconsistent with $\psi$ being in $\HH_0 \otimes \FF$. Thus
$H(\xi)$ does not have a ground state. \qed

\section*{Acknowledgements}

D.H. wants to thank Marcel Griesemer, Volker Bach, and Michael Loss
for  interesting discussions. I.H. would like to acknowledge an
interesting conversation with Benoit Gr\'{e}bert.

\end{document}